\documentclass{aa}  

\usepackage{graphicx}
\usepackage{subfig}
\usepackage{hyperref}
\usepackage{txfonts}
\usepackage{float}
\usepackage{amssymb}
\begin{document}

   \title{Nonthermal radiation from the central region of super-accreting active galactic nuclei}
	
	\titlerunning{Nonthermal emission from super-accreting AGNs}

   \author{Pablo Sotomayor\inst{1,2}\and
           Gustavo E. Romero\inst{1,2}
          }

   \institute{Instituto Argentino de Radioastronom\'{\i}a (CCT-La Plata, CONICET; CICPBA; UNLP), C.C. No. 5, 1894,Villa Elisa, Argentina\\
              \email{psotomayor@iar.unlp.edu.ar}
         \and
Facultad de Ciencias Astron\'omicas y Geof\'{\i}sicas, Universidad Nacional de La Plata, Paseo del Bosque s/n, 1900, La Plata, Argentina
        }

   \date{Received ***; accepted ***}

 
  \abstract
  {The radio emission mechanism in active galactic nuclei (AGN) with high accretion rates is unclear. It has been suggested that low-power jets may explain the observed radiation at sub-parsec scales. The mechanisms for jet formation at super-Eddington rates, however, are not well understood. On the same scale, clouds from the broad-line region (BLR) propagating with supersonic velocities in the wind launched by the accretion disk may lead to the production of nonthermal radiation.}
   {We aim to characterize the nonthermal emission produced by the propagation of clouds through the wind of the accretion disk in super-accreting AGNs, and to estimate the relevance of such contribution to the radio band of the electromagnetic spectrum.}
   {We determine the conditions under which the BLR clouds are not destroyed by shocks or hydrodynamic instabilities when immersed in the powerful wind of the accretion disk. These clouds form bowshocks that are suitable sites for particle acceleration. We develop a semianalytical model to calculate the distribution of relativistic particles in these bowshocks and the associated spectral energy distribution (SED) of the emitted radiation.}
   {For typical parameters of super-accreting AGNs, we find that the cloud-wind interactions can produce nonthermal emission from radio up to a few tens of TeV, with slight absorption effects, if the processes occur outside the wind photosphere.}
   {Radio emission in AGNs without jets can be explained if the accretion rate is super-Eddington and there is a broad-line region at sub-parsec scales around the central black hole. The accretion rate must not be extremely high so most of the clouds orbit outside of the wind photosphere and the radiation can escape to the observer. Instabilities in the disk wind previously reported in numerical simulations, generate clumps that increase the filling factor of the overdensities in the broad-line region and enhanced the emitted radiation.}
   
   
   \maketitle
%

\section{Introduction}
\label{sect:intro}
Accretion of matter onto a black hole proceeds in three basic regimes. If we denote the Eddington luminosity\footnote{$  L_{\rm Edd}=4 \pi c G M m_{\rm p}  \sigma _{\rm T}^{-1}\approx 1.3\times10^{38} (M/M_{\odot})\;{\rm erg\, s^{-1}}$ is the luminosity required to stop the accretion process through radiation pressure in spherical accretion.} by $L_{\rm Edd}$ and  we define a critical accretion rate as $ \dot{M}_{\rm cr} \equiv \eta \dot{M}_{\rm Edd} \approx 1.4\times 10^{17} (M/M_{\odot})\;\; {\rm g \;s}^{-1}$, where $\eta\sim0.1$ is the accretion efficiency, then at subcritical rates $\dot{M}<\dot{M}_{\rm cr}$ the accretion flow forms a geometrically thin and optically thick disk that can be described by the standard $\alpha$-disk solution found by \cite{shakura1973}. At very low rates, $\dot{M}\ll \dot{M}_{\rm cr}$, the inner disk inflates and becomes optically thin. The gas is advected before it can cool, and becomes very hot with temperatures $T_{\rm e}\sim10^9$ K for electrons and $T_{\rm p}\sim (m_{\rm p}/m_{\rm e}) \; T_{\rm e} \sim 10^{12}$ K for protons. This regime, usually called Advection Dominated Accretion Flow (ADAF), can be described by the self-similar solutions given by \cite{Ichimaru1977} and \cite{narayan1994}. The ADAF region is similar to a two-temperature corona surrounding the black hole  \citep[e.g.,][]{Bisnovatyi-Kogan1976,Liang-Price1977}. This hot plasma Comptonizes the softer radiation from the disk, producing the hard X-ray emission observed up to photon energies of $\sim 150$ keV that characterizes the spectrum of many X-ray binaries.\par
When the accretion regime is super-critical, i.e.  $\dot{M}>\dot{M}_{\rm cr}$,  photon trapping in the disk becomes important, the disk inflates becoming thick or slim, and the gas is advected to the black hole. In this regime the radiation pressure dominates the innermost region and most of the accreting matter in excess of the Eddington rate is evacuated through a wind \citep[e.g.,][]{shakura1973,lipunova1999,fukue2004}. The luminosity sets around the Eddington value. Super-critical disks are thought to feed powerful microquasars such as SS433, some active galactic nuclei, transient episodes of accretion associated with tidal disruption of stars in supermassive black holes, and ultraluminous X-ray sources observed in other galaxies.\par
Semi-analytical models for supermassive black holes with high accretion rates are usually applied to Narrow-line Seryfert 1 (NLS1) galaxies \citep[see e.g.][]{fukue2004}. NLS1 galaxies are low-mass AGNs that constitute about 15\% of Seyfert 1 galaxies and quasars at cosmological redshifts $z < 0.5$ \citep{williams2002}. From the relation between the size of the broad-line region and the optical luminosity in AGNs derived by \cite{kaspi2000}, it follows that NLS1 galaxies have super-Eddington accretion rates and high Eddington ratios ($\lambda_{\rm Edd}\equiv L_{\rm bol}/L_{\rm Edd}$) \citep[see e.g.][]{collin2004,peterson2011}. Using observations with the Very Large Array, \cite{greene2006} found that NLS1 galaxies with high Eddington ratio are generally radio quiet. In these systems, the jets, if present at all, are faint. Several alternative radio emission mechanisms have been proposed for these sources: star formation, AGN winds, or coronal activity \cite[see][]{panessa2019}. AGN winds can explain radio emission on scales of a few hundred parsecs through synchrotron radiation of electrons accelerated by wind shocks \citep{nims2015}. \par
Several high-resolution radio interferometric surveys of the nuclear region of NLS1 galaxies  \citep[e.g.][]{Berton2018,Chen2020,Javerla2022} reveal many sources with a compact radio structure and a steep spectral index. This is contrary to the hypothesis that the radio emission is originated from a compact jet, which should manifest as an elongated structure with a flat spectral index. The origin of this nuclear radio emission in otherwise radio silent galaxies demands an explanation consistent with the accretion regime of the sources \citep[see a relevant discussion in][]{panessa2019}. In the present paper we propose that in radio-quiet galaxies with super-Eddington accretion rates, shock waves associated with bowshocks produced by wind-cloud interactions can accelerate particles and generate nonthermal radiation, and in particular the nuclear synchrotron radiation observed in many sources. 

Nonthermal radiation caused by the propagation of clouds into outflows has been previously investigated by \cite{araudo2010} and \cite{delpalacio2019} in the context of AGN jets.  \cite{muller2020a} studied the effects of the impact of BRL clouds onto AGN disks. The interaction of clouds and galactic winds in starburst galaxies can also give rise to a broadband nonthermal spectrum and has been studied by \cite{muller2020b}. \cite{moriya2017} investigated the production of thermal radiation in AGNs with high accretion rates when strong radiative shocks propagate through the BLR. What we present in this work is the first investigation of the production of nonthermal photons by the interaction of clouds with the wind of the disk in super-accreting AGNs. 

The propagation of BLR clouds in the wind results in the generation of shock waves that, under adiabatic conditions, accelerate charged particles up to relativistic energies. The maximum energy is limited by the cooling processes, the escape of particles, and the size of the acceleration region. We adopt a stationary one-zone model for the treatment of particle transport and calculate the nonthermal radiation because of synchrotron mechanism, inverse Compton upscattering of reprocessed disk photons, and relativistic Bremsstrahlung for the case of electrons, plus synchrotron and neutral pion decay for protons.\par

This paper is organized as follows: in Section \ref{sect:model} we present the model and discuss its basic components: winds from super-critical accretion disks, dynamics of wind-cloud interactions, and nonthermal radiative processes. We apply our model first to a generic case and present the results in Section \ref{sect:results}. Then, in Section \ref{sect:application} we present an application to a specific case of a NLS1 galaxy, Mkr 335, followed with the discussion and conclusions in Sections \ref{sect:discussion} and \ref{sect:conclusions}. We adopt cgs units throughout the paper.

\section{Model}
\label{sect:model}

\subsection{Disk and winds}
\label{subsect:winds}

When the accretion regime is super-critical, the disk is divided into two regions: an outer standard disk without winds, and an inner, radiation-dominated disk that accretes matter at the Eddington rate and ejects the excess of gas in the form of a radiation-driven wind \citep[see e.g.][]{fukue2004}. Both regions are separated at the critical radius $r_{\rm cr}$, where the vertical component of the gravitational acceleration is balanced by the radiation flux of the disk. This radius is given by \citep{fukue2004}:
\begin{equation}
r_{\rm cr}=\frac{9\sqrt{3}\sigma_{\rm T}}{16\pi m_{\rm p}c}\dot{M} \approx 4\dot{m}r_{\rm g},
\end{equation}
where $\sigma_{\rm T}$ is the Thomson cross section, $m_{\rm p}$ the proton mass, $c$ the speed of light, $\dot{m} = \dot{M}/\dot{M}_{\rm cr}$ the normalized accretion rate, and $r_{\rm g}$ is the gravitational radius of the black hole. 

In the inner region of the disk the strong mass loss forms a wind driven by radiation. Although the accretion rate is super-Eddington, the luminosity of the disk is of the order of the Eddington limit or slightly greater. The bolometric luminosity of super-accreting disks can be estimated as \citep{fukue2004}
\begin{equation}
\label{eq:disk luminosity}
L_{\rm bol} = L_{\rm Edd} \frac{2}{3\sqrt{3}} \left(1 + \ln \left(4 \dot{m} \frac{r_{\rm g}}{r_{\rm in}}\right) \right).
\end{equation}

The wind launched from $r < r_{\rm cr}$ is optically thick, spherically symmetric, and mildly relativistic wind \citep[e.g.][]{fukue2009,fukue2010}. We consider a steady state in such a way that $\dot{M}_{\rm w}$, $v_{\rm w}$, $L_{\rm w}$ are constants, where $\dot{M}_{\rm w}$ is the mass-outflow rate, $v_{\rm w}$ is the wind terminal velocity, and $L_{\rm w}$ is the wind bolometric luminosity in the comoving frame. The scaling relation between the terminal velocity and the escape velocity for line-driven winds from early-type stars is given by \citep[see Eq.~(13) in][]{puls2008}:
\begin{equation}
v_{\rm w} \approx 2.25 \frac{\alpha}{1 - \alpha}v_{\rm esc},
\end{equation}
where $\alpha \approx 0.8$ for O stars. Here, we consider the escape velocity at the critical radius of the disk, where the dynamics becomes radiation-dominated. The velocity should then be corrected by Thomson acceleration, that is, it takes into account the radiation dragging. Hence, $v_{\rm esc} < \sqrt{2GM_{\rm BH}/r_{\rm cr}}$. The corrected value is \citep{king2010}:
\begin{equation}
\label{eq:wind velocity}
v_{\rm w} = \sqrt{\frac{GM_{\rm BH}}{r_{\rm cr}}} = \frac{c}{2\,\sqrt{\dot{m}}},
\end{equation}
where $M_{\rm BH}$ is the black hole mass and $\dot{M}_{\rm w} \approx \dot{M}$, which is a adequate for super-accreting sources, and is used throughout the paper.\par

The apparent photosphere of the wind is defined as the surface where the optical depth $\tau_{\rm ph}$ is unity for an observer at infinity. The wind velocity might be relativistic, so the optical depth in the observer frame depends in general on the magnitude of the velocity and the viewing angle. The location of the apparent photosphere from the equatorial plane $z_{\rm ph}$ is given by \citep[][]{fukue2010}:
\begin{equation}
\tau_{\rm ph} = - \int	_{\infty} ^{z_{\rm ph}} \Gamma_{\rm w} \left(1 - \beta \cos \theta \right)\kappa_{\rm co}\rho_{\rm co}{\rm d}z = 1,
\label{eq:apparent photosphere}
\end{equation} 
where $\Gamma_{\rm w}$ the wind Lorentz factor, $\beta = v_{\rm w}/c$ is the wind normalized velocity, $\theta$ the angle of the flow with respect to the line-of-sight, $\kappa_{\rm co}$ the opacity in the comoving frame, and $\rho_{\rm co}$ the wind density in the comoving frame. We assume a highly ionized wind such that the opacity is dominated by free electron scattering ($\kappa_{\rm co} = \sigma_{\rm T}/m_{\rm p}$).\par

In the region inside the wind photosphere, radiation is absorbed and cannot escape. Since the main goal of our work is to estimate the nonthermal radiation emitted by BLR clouds within the wind that reaches the observer, the clouds we consider orbit outside the photosphere. This imposes a restriction on our model as we show in Section \ref{sect:results}: the accretion rate should not be extremely super-Eddington; otherwise most clouds would be inside the opaque region. 


\begin{figure}[h!]
 \centering 
 \includegraphics[scale=0.31]{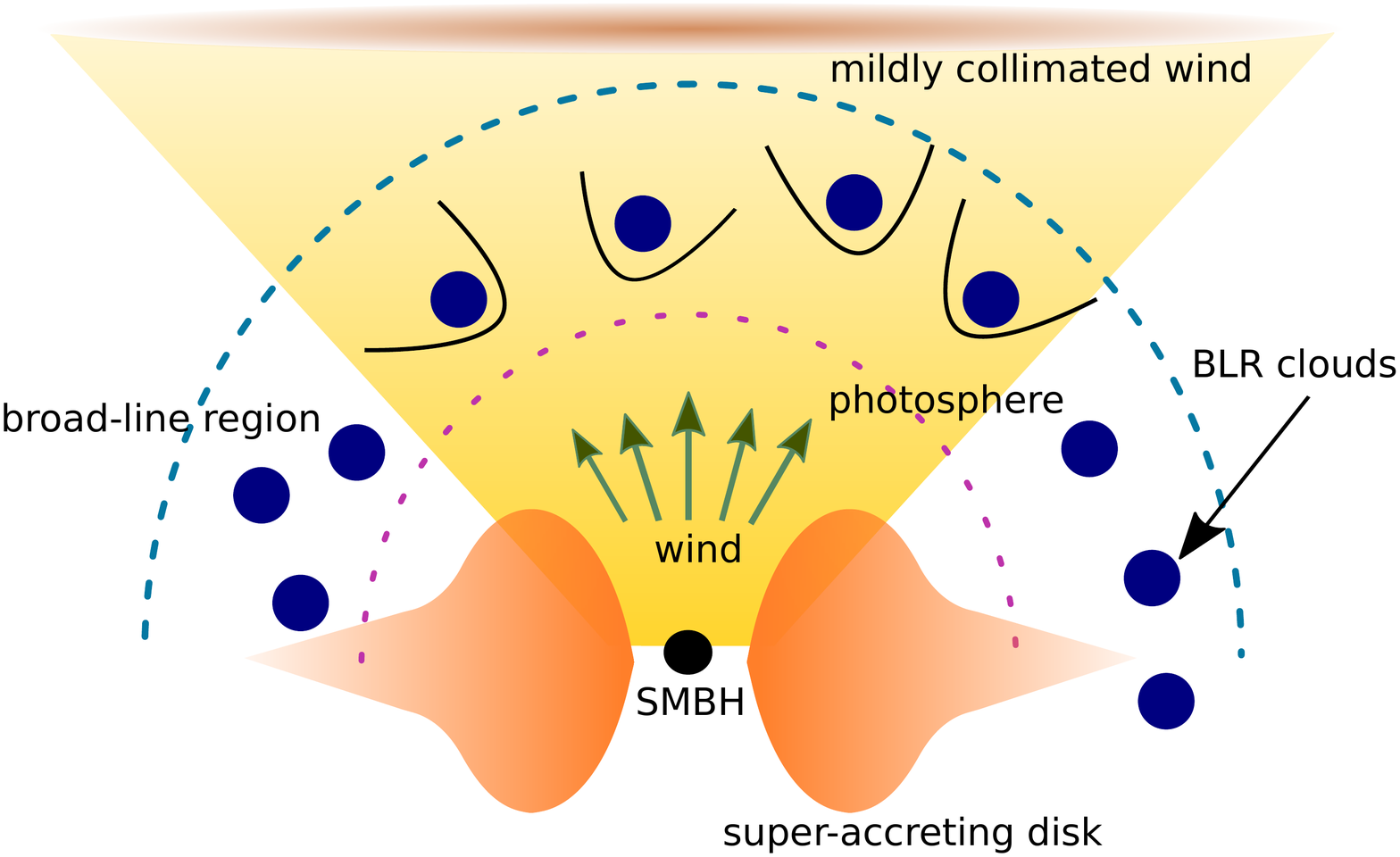}
 \caption{\small Sketch of BLR clouds propagating in the wind of a super-accreting black hole (not to scale). The clouds move in Keplerian orbits around the black hole with the velocity vectors being isotropically distributed in the tangent plane defined by $r = {\rm constant}$, with $r$ the distance from the cloud to the black hole. Bowshocks only form around clouds moving within the wind.}
 \label{fig:scheme}
\end{figure}

\begin{figure}[h!]
 \centering 
 \includegraphics[scale=0.31]{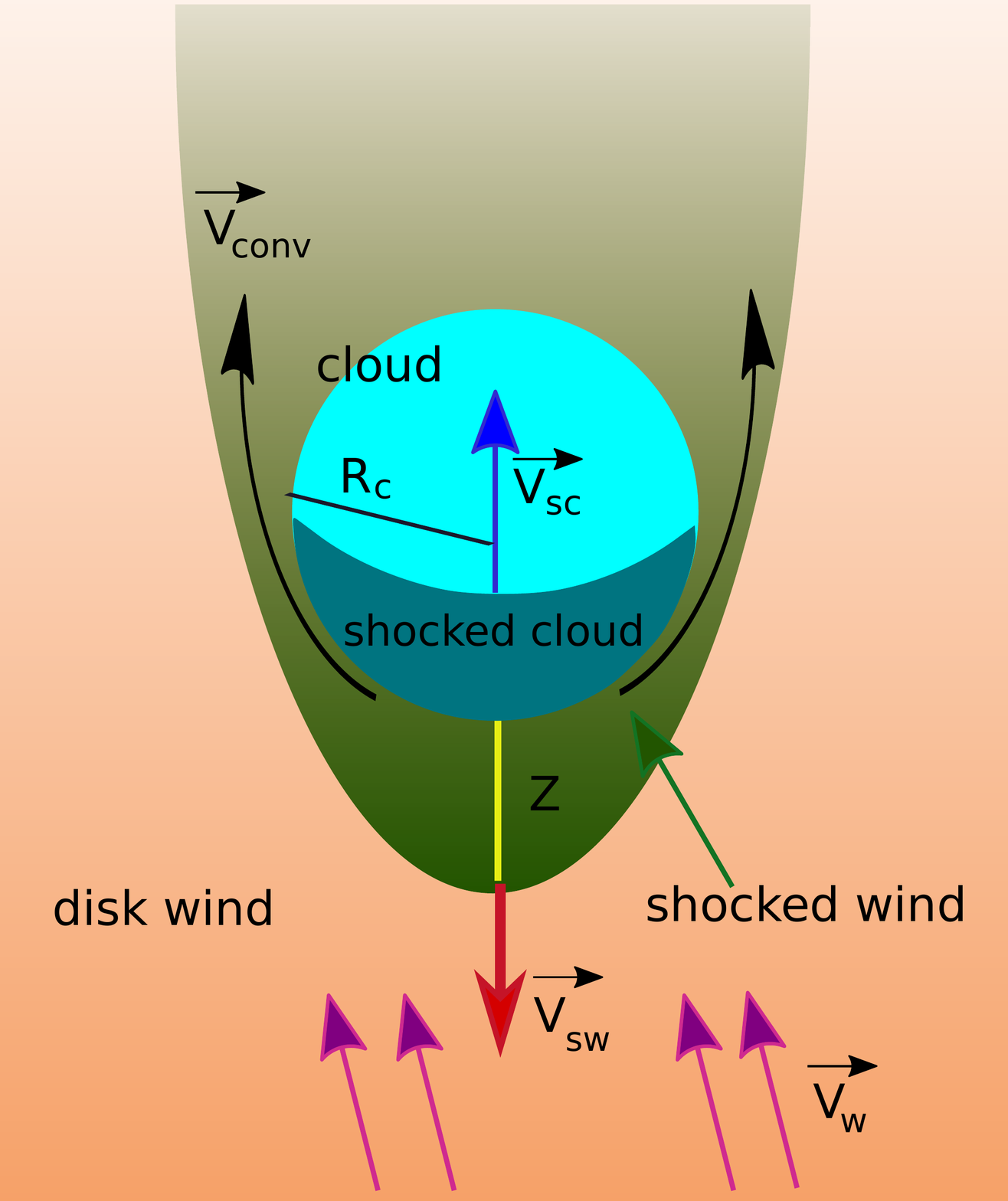}
 \caption{\small Representation of the interaction between the wind of the accretion disk and a cloud. The shock in the cloud propagates with velocity $v_{\rm sc}$, and the shock in the wind with velocity $v_{\rm sw}$. Velocities of the wind and shocks are not aligned because the semi-major axis of the bowshock is directed along the composition of the cloud and wind velocities.}
 \label{fig:scheme zoom}
\end{figure}
\subsection{Wind-cloud interactions}
\label{subsect:interactions}


In order to estimate the mass of quasars, reverberation techniques are used. This procedure relates the size of the broad-line region, the velocity of the clouds, and the mass of the central black hole \citep[see][]{kaspi2000}. One important assumption is that the dynamics of BLR clouds is dominated by the gravitational force of the black hole. The typical velocity of the clouds can be estimated with this technique as:
\begin{equation}
\label{eq:vc}
v_{\rm c}^2 = f^{-1} \frac{GM_{\rm BH}}{R_{\rm BLR}},
\end{equation}
where $R_{\rm BLR}$ is the radius of the BLR, and the factor $f = 3/4$ corrects three-dimensional effects of cloud motion \citep[see Eq.(5) in][]{kaspi2000}. This factor is necessary because the velocity included in the reverberation mapping is obtained from the spectrum of the gas orbiting the central black hole, so it is a projected magnitude.\par 

If $N_{\rm c}$ is the total number of clouds in the BLR, the number of clouds within the wind $N_{\rm c}^{\rm w}$ is given by
\begin{equation}
\label{eq:clouds inside the wind}
N_{\rm c}^{\rm w} = 2\frac{\Omega_{\rm w}}{4\pi} N_{\rm c},
\end{equation}
where the factor $2$ accounts for the two outflows from both sides of the disk, and $\Omega_{\rm w}$ is the wind solid angle. The filling factor of the clouds in the broad-line region is
\begin{equation}
\label{eq:filling factor}
f_{\rm BLR} = N_{\rm c} \frac{V_{\rm c}}{V_{\rm BLR}},
\end{equation}
where $V_{\rm c}$ is the volume of a single cloud, and $V_{\rm BLR} = 4\pi(R_{\rm BLR}^3 - R_{\rm in}^3)/3$ is the volume of the BLR. Typically, $R_{\rm in} = (0.6 - 0.8)\,R_{\rm BLR}$ \citep[see][]{savic2020}. 

When a cloud enters into the wind, two shocks are generated: a shock in the cloud and a shock in the wind. We assume that the clouds describe circular orbits around the supermassive black hole, so the velocity vector of each cloud is  perpendicular to the velocity of the wind, which is radial. 

The semimajor axis of the bowshock formed in the wind is directed along the composition of the velocity vectors of the cloud and the wind, similar to the case of bowshocks generated by the motion of a compact object (e.g. a neutron star) in the wind of a massive star \citep[see e.g. the Chapter 4 in][]{frank1992}. Shock velocities are given by \citep[see e.g.][and references therein]{muller2020b}
\begin{equation}
\label{eq:veloc sc}
v_{\rm sc} = \frac{4}{3}\frac{\sqrt{v_{\rm w}^2 + v_{\rm c}^2}}{1 + \sqrt{n_{\rm c} / n_{\rm w}}},
\end{equation}
and
\begin{equation}
\label{eq:veloc sw}
v_{\rm sw} = \frac{4}{3}\frac{\sqrt{v_{\rm w}^2 + v_{\rm c}^2}}{1 + \sqrt{n_{\rm w} / n_{\rm c}}},
\end{equation}
where $v_{\rm sc},\;v_{\rm sw}$ are the velocities of shock in the cloud and shock in the wind respectively, and $n_{\rm w}$ is the wind density, which is obtained from the continuity equation for the wind.\par

Diffusive shock acceleration (DSA) of charged particles occurs when the shock is adiabatic. If, on the contrary, the shock is radiative, the shocked region cools very quickly and the large increase in entropy destroys the inhomogeneities in the magnetic field on both sides of the shock; these magnetic structures are necessary for particle acceleration. In order to check for adiabaticity, we calculate the typical lengthscale of radiative cooling $R_{\Lambda}$ for the shocked regions. This lengthscale is the relaxation length for which the gas reaches thermal equilibrium because of the radiative losses \citep{mccray1979}:
\begin{equation}
\label{eq:cooling lengthscale}
R_{\Lambda} = \frac{1.14\times10^{-29}\;(v_{\rm sh} / {\rm km\,s^{-1}})^3}{\left(n / {\rm cm^{-3}}\right) \left(\Lambda(T) / {\rm erg\,cm^{3}\,s^{-1}} \right)}\;{\rm pc},
\end{equation}
where
\begin{equation}
\label{eq:shock temperatures}
T = 10.9 \left(\frac{v_{\rm sh}}{\rm km \,s^{-1}} \right)^2\;{\rm K}.
\end{equation}

If $R_{\Lambda}$ is significantly greater than the lengthscale of the shocked regions, the shocks are adiabatic; otherwise they are radiative. The thermal emissivity in the shocked medium depends on the dominant opacity mechanism and, therefore, on the gas temperature \citep[see][]{wolfire2003}:
\begin{equation}
\label{eq:shock emissivities}
\Lambda(T) =
\begin{cases}
7\times 10^{-27} T,\quad 10^{4}\, {\rm K} \leq T \leq 10^{5} {\rm K}\\
7\times 10^{-19} T^{-0.6},\quad 10^{5}\, {\rm K} \leq T \leq 4\times 10^{7} {\rm K}\\
3\times 10^{-27}\, T^{0.5},\quad T \geq 4\times 10^{7} {\rm K}
\end{cases}
\end{equation}



The time required for a cloud of radius $R_{\rm c}$ and velocity $v_{\rm c}$ to fully enter the wind is given by (see Fig.~\ref{fig:scheme}):
\begin{equation}
t_{\rm c} = \frac{2\,R_{\rm c}}{v_{\rm c}}.
\end{equation}
Clouds fully penetrate the wind if $t_{\rm c}$ is shorter than the timescale for the clouds to be destroyed by 1) hydrodynamic instabilities, and 2) shocks in the clouds. The crossing time of the shock in the cloud is \begin{equation}
t_{\rm crush} = \frac{2\,R_{\rm c}}{v_{\rm sc}}.
\end{equation}

The relative motion of fluids produces hydrodynamic instabilities. The density contrast between the cloud and the wind leads to Rayleigh-Taylor instabilities at the frontal interface between them. The acceleration exerted by the wind on the surface of the cloud is given by
\begin{equation}
a_{\rm c} = \frac{P_{\rm ram,\,w} \pi R_{\rm c}^2}{M_{\rm c}},
\end{equation}
where $P_{\rm ram, \,w} = n_{\rm w}\mu m_{\rm p}v_{\rm w}^2$ is the ram pressure of the wind, where $\mu$ is the mean molecular weight of the gas (we adopt $\mu = 0.6$, corresponding to an ionized plasma). The timescale for Rayleigh-Taylor instabilities to develop at the fluid interface is \citep[][]{araudo2010}:
\begin{equation}
t_{\rm RT} = \sqrt{\frac{2\,R_{\rm c}}{a_{\rm c}}}.
\end{equation}

Kelvin-Helmholtz instabilities also manifest at the fluid interface. The timescale for these instabilities is \citep{muller2020b}
\begin{equation}
t_{\rm KH} = \frac{2\,R_{\rm c}\,(n_{\rm c} + n_{\rm w})}{(v_{\rm w} - v_{\rm c})\sqrt{n_{\rm c}\,n_{\rm w}}}.
\end{equation}

The typical time for the clouds to be destroyed by shocks or instabilities limits the time of production of nonthermal radiation in the wind-cloud interactions\footnote{We note that the momentum transferred from the wind can move the clouds away (but always with the cloud remaining within the wind, since the applied force is radial), taking them to regions where the timescale of destruction by instabilities will be longer. A correct quantification of this effect requires hydrodynamic simulations and is beyond the scope of the present paper.}. According to Eqs. (\ref{eq:cooling lengthscale} - \ref{eq:shock emissivities}), shocks propagating through the clouds are radiative \citep[see also][]{moriya2017} and the shocks propagating in the wind are typically adiabatic. Acceleration of particles takes place mainly in these latter adiabatic shocks.\par 

In Fig. \ref{fig:scheme} we show the overall picture of our model. A more detailed sketch of the interaction between the wind and a single cloud is presented in Fig. \ref{fig:scheme zoom}.

\subsection{Particle acceleration and radiative processes}
\label{subsect:radiation}
The emergent nonthermal spectrum from the many bowshocks formed in the system depends on several factors: the efficiency of transferring energy to relativistic particles, the hadron-to-lepton power ratio, the dominant cooling mechanisms, the spectral index of the power law for the injection of particles, and absorption effects.\par

The kinetic power of the shock in the wind is given by
\begin{equation}
\label{eq:bowshock kinetic power}
L_{\rm k,bs} = \frac{1}{2}n_{\rm w}m_{\rm p}v_{\rm sw}^3 A_{\rm bs},
\end{equation}
where $A_{\rm bs}$ is the area of the shocked region, $A_{\rm bs} = 2\pi (R_{\rm c} + Z)^2$, with $Z$ being the distance from the cloud to the contact discontinuity between the bowshock and the unshocked wind, in the direction of the bowshock apex.\par 

We assume that a small fraction of the total kinetic power of the wind $q_{\rm rel}$ is transferred into relativistic particles by diffusive shock acceleration:
\begin{equation}
\label{eq:relativistic particles power}
L_{\rm rel} = q_{\rm rel} L_{\rm k,bs},
\end{equation}
where $L_{\rm rel}$ is distributed between protons and electrons:
\begin{equation}
L_{\rm p} = a L_{\rm e},
\end{equation}
where $a$ is the hadron-to-lepton power ratio. This is a free parameter and we explore different values in this work.\par
The injection function is a power-law in the particle energy with an exponential cut-off \citep[see e.g.][]{drury1983},
\begin{equation}
\label{eq:injection function}
Q(E) = Q_{0}\, E^{-p} \,\exp{\left(-E/E_{\rm max}\right)},
\end{equation}
where $Q_0$ is a normalization constant (different for each particle species), and $p$ is the particle injection index. We assume that electrons and protons are injected with the same spectral index.The normalization constant $Q_{0}$ is obtained from
\begin{equation}
L_{(\mathrm{e,p})} = \Delta V \int ^{E_{(\mathrm{e,p})} ^{\mathrm{max}}} _{E_{(\mathrm{e,p})}^{\mathrm{min}}}\mathrm{d}E_{(\mathrm{e,p})}E_{(\mathrm{e,p})}Q_{(\mathrm{e,p})}(E_{(\mathrm{e,p})}),
\label{eq: particle luminosity}
\end{equation}
where $\Delta V$ is the volume of the acceleration region, and $E_{\rm e,\,p}^{\rm max}$ is the maximum energy reached by these particles. The latter is obtained by balancing the total loss time (radiative and nonradiative) of the particles with the acceleration time. Formulas for calculating the timescales of cooling, acceleration, and particle escape are provided in Appendix~\ref{app1}.\par

We obtain the distribution of relativistic particles $N\,[{\rm erg^{-1}\;cm^{-3}}]$ solving the transport equation in phase space. The transport equation is given by \citep{Ginzburg1964}:
\begin{equation}
\frac{\partial N}{\partial t} - \nabla \cdot (D\nabla N) + \frac{\partial }{\partial E}\left[\left.\frac{\mathrm{d}E}{\mathrm{d}t}\right\vert_{\mathrm{loss}}N\right] + \frac{N}{t_\mathrm{esc}} = Q(E,t),
\end{equation}
where $D$ is the diffusion coefficient. Since the timescale for the bowshock formation is lower than the timescale of the cloud destruction processes (see Fig.~\ref{fig:shocks timescales}), we can assume that the bowshock is in a steady state. Spatial variations in fluid dynamic variables within bowshocks are small and can be neglected, so we adopt a simple stationary one-zone model such that this equation takes the simplified form:
\begin{equation}
\label{eq:particle transport}
\frac{\partial }{\partial E} \left[\left.\frac{\mathrm{d}E}{\mathrm{d}t}\right\vert_{\mathrm{loss}}N(E)\right] + \frac{N(E)}{t_\mathrm{esc}} = Q(E),
\end{equation}
where the $t_{\rm esc}$ is the timescale for the particle escape of the acceleration region. We consider particle escape by convection and diffusion throughout this paper.\par

Finally, we calculate the spectral energy distribution (SED) of the radiation produced by the relativistic particles in their interaction with magnetic, matter and radiation fields in the wind. Detailed formulae for the emissivities of each radiative process can be found in \cite{reynoso2009}, \cite{romero2010a}, \cite{vieyro2012}, \cite{vila2012PhDT}, \cite{romero2014}, and references therein. The total luminosity for any of the leptonic/hadronic radiative processes is calculated as $E_{\gamma}\,L_{\gamma}(E_{\gamma})$ (in units of ${\rm erg\,s^{-1}}$), where $L_{\gamma}(E_{\gamma})$ is the specific luminosity (in units of ${\rm erg\,s^{-1}\,erg^{-1}}$) at energy $E_{\gamma}$. Similarly, the total flux is calculated as $E_{\gamma}\,F_{\gamma}(E_{\gamma})$ (in units of ${\rm erg\,s^{-1}\,cm^{-2}}$), where the specific flux $F_{\gamma}(E_{\gamma})$ is in units of ${\rm erg\,s^{-1}\,cm^{-2}\,erg^{-1}}$.\par

The emerging spectrum should include corrections for absorption. At very-high energies, the relevant absorption mechanism is $\gamma \gamma$-annihilation. We consider two fields of photons for this absorption: photons from the wind photosphere (external absorption), and photons from the electron synchrotron radiation field (internal absorption). In addition, we correct the emitted radio luminosity by synchrotron self-absorption.\par

\section{Results}
\label{sect:results}

\subsection{Wind photosphere}
\label{subsect:wind}

In this section we apply our model to a generic AGN scenario. We consider a supermassive black hole of $M_{\rm BH} = 10^7\,M_{\odot}$, surrounded by a broad-line region of size $R_{\rm BLR} = 10^4\,r_{\rm g} = 1.5\times10^{16}\,{\rm cm}$. The accretion rate determines the location of the wind photosphere. The radiative power of the wind in super-critical sources is limited to the Eddington luminosity (see \cite{meier1982} for a theoretical derivation from the radiation hydrodynamic equations, and \cite{zhou2019} for an observational estimate), henceforth we adopt a value $L_{\rm w} = 0.1\, L_{\rm Edd}$.\par

\begin{figure}[h!]
 \centering 
 \includegraphics[scale=0.50]{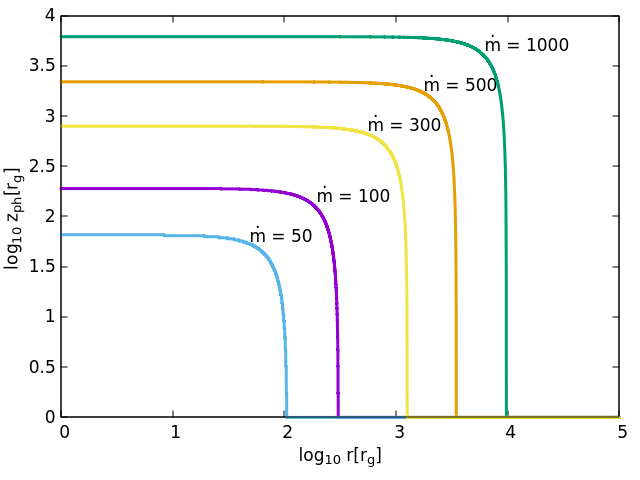}
 \caption{\small Location of the apparent photosphere of the wind for several values of the accretion rate. The mass of the black hole is $M_{\rm BH} = 10^7\,M_{\odot}$, and the wind luminosity is $L_{\rm w} = 0.1\,L_{\rm Edd}$.}
 \label{fig:apparent photosphere}
\end{figure}

In Fig. \ref{fig:apparent photosphere} we show the solutions to equation (\ref{eq:apparent photosphere}) for $\dot{m} = 50,\, 100 ,\, 300,\, 500\, {\rm and}\, 1000$. The wind velocity corresponding to these accretion rates is calculated using Eq. (\ref{eq:wind velocity}), yielding $v_{\rm w}/c = 0.071,\,0.050,\,0.037,\,0.022,\,{\rm and }\,0.016$, respectively. Even for the hyper-accreting case, $\dot{m} = 10^3$, the apparent photosphere of the wind is located at sub-parsec scales of $z_{\rm ph} \approx 10^4\,r_{\rm g} \approx 5\times10^{-3}\,{\rm pc}$. From now on we adopt $\dot{M}_{\rm w} = 100\,\dot{M}_{\rm cr}$, so the radiation from all bowshocks can escape to the observer. \par

\begin{figure}[h!]
 \centering 
 \includegraphics[scale=0.50]{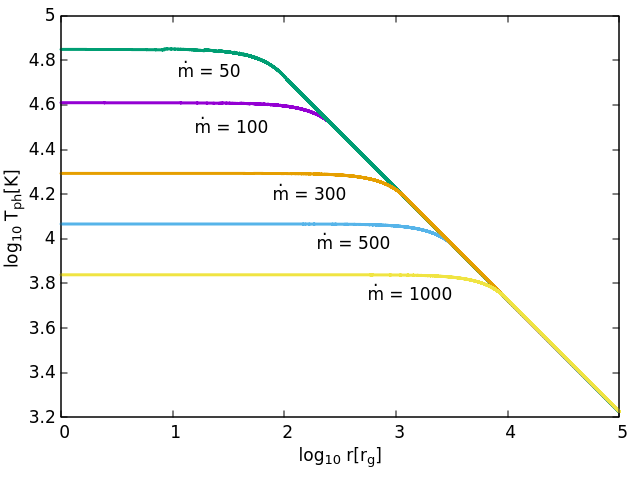}
 \caption{\small Radial distribution of the temperature at the apparent photosphere for several values of accretion rate. As before, $M_{\rm BH} = 10^7\,M_{\odot}$, and $L_{\rm w} = 0.1\,L_{\rm Edd}$.}
 \label{fig:wind temperature}
\end{figure}

\begin{figure}[h!]
 \centering 
 \includegraphics[scale=0.50]{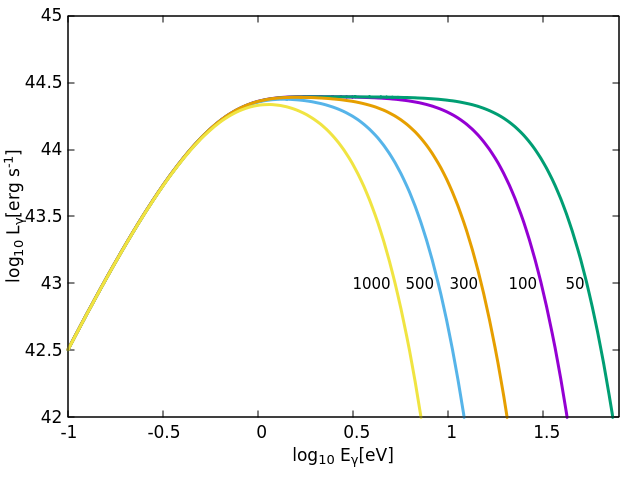}
 \caption{\small Spectral energy distribution of the emitted radiation at the apparent photosphere for several values of accretion rate. As before, $M_{\rm BH} = 10^7\,M_{\odot}$, and $L_{\rm w} = 0.1\,L_{\rm Edd}$. The numbers next to each curve indicate the accretion rate in critical accretion rate units.}
 \label{fig:wind SED}
\end{figure}

Once the location of the photosphere has been determined, we obtain the temperature distribution through the wind in the comoving frame assuming that the gas emits locally as a blackbody. 
Figures \ref{fig:wind temperature} and \ref{fig:wind SED} show this temperature and the observed wind spectrum in the comoving frame. The higher the accretion rate, the lower the temperature is in the apparent photosphere. This is expected since the temperature goes as $T \propto r^{-2}$. Accordingly, the characteristic energy of the photons from the photosphere is higher for lower accretion rates.\par
\begin{table}[h]
\centering                          
\caption{\small Parameters for the broad-line region and clouds}             
\label{table:clouds}      

\begin{tabular}{l c} 
\hline 
Parameter & Value  \\
\hline       
   Black hole mass & $M_{\rm BH} = 10^7\,{M}_{\odot}$ \\
   Wind mass-loss rate & $\dot{M} = 100\,\dot{M}_{\rm cr}$ \\
   Wind solid angle& $\Omega_{\rm w} = \pi\,{\rm sr}$\\
   Wind velocity$^\dagger$& $v_{\rm w} = 0.05\,c$\\
   Broad-line region radius$^\ddagger$& $R_{\rm BLR} = 1.5\times10^{16}\,{\rm cm}$\\
   Broad-line region filling factor$^\Box$ & $f_{\rm BLR} = 10^{-6}$\\
   Cloud size$^\diamond$ & $R_{\rm c} = 3.2\times10^{11}\,{\rm cm}$ \\
   Cloud density$^\diamond$ & $n_{\rm c} = 10^{10}\,{\rm cm^{-3}}$ \\   
   Cloud velocity$^\nabla$ & $v_{\rm c} = 3.5\times10^{8}\,{\rm cm\,s^{-1}}$ \\   
   Bowshock-cloud separation$^\oplus$ & $Z = 0.3\,R_{\rm c}$ \\
   \hline    
$^{\dagger}$: \small  \cite{king2010}. \\
$^{\ddagger}$: \small  \cite{kaspi2000}.\\
$^{\Box}$: \small  \cite{netzer1990}.\\ 
$^{\diamond}$: \small  \cite{risaliti2010}.\\
$^{\nabla}$: \small  \cite{peterson2006}.\\
$^{\oplus}$: \small  \cite{araudo2010}.
\end{tabular}
\end{table}

\subsection{About the clouds}
\label{subsect:dynamics}


We adopt the parameters listed in Table \ref{table:clouds} to characterize the clouds of the BLR. To estimate the size of the clouds we use a filling factor in the broad-line region of $f_{\rm BLR} = 10^{-6}$ \citep{netzer1990}. We assume that the total number of clouds is $N_{\rm c} = 10^8$ (this value is typically assumed in the literature, see e.g. \citet{arav1997,dietrich1999}, and much smaller than the maximum value $N_{\rm c} = 10^{18}$ found by \cite{abolmasov2017}), and their particle number density is $n_{\rm c} = 10^{10}\,{\rm cm^{-3}}$ \citep{risaliti2010}. In addition, for sake the of simplicity, we assume a population of identical spherical clouds.

\begin{figure}[h!]
 \centering 
 \includegraphics[scale=0.60]{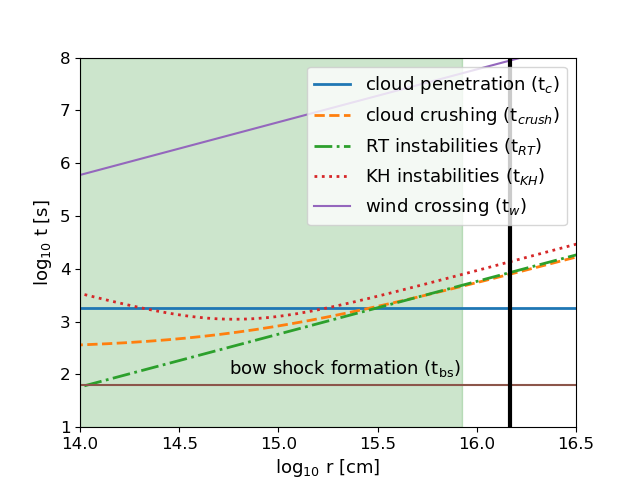}
 \caption{\small The different timescales for processes affecting a cloud during its interaction with the wind of the accretion disk. The thick black vertical line indicates the position of the broad-line region. The shaded region corresponds to distances opaque to radiation.}
 \label{fig:shocks timescales}
\end{figure}

Figure \ref{fig:shocks timescales} shows the timescales relevant for the clouds to survive to the interactions with the wind, estimated using the formulae presented in Sect. \ref{subsect:interactions}. The clouds can completely penetrate the wind if they are located at $r\geqslant 8.4\times10^{15}\,{\rm cm}$, so essentially all clouds in orbits that intersect the wind get inside it. The timescale to cross the wind region is $t_{\rm w}$. Since ${\rm max}\left(t_{\rm crush},  t_{\rm KH}, t_{\rm RT}\right) < t_{\rm w}$, all clouds are ablated in the wind\footnote{The full destruction of the clouds can be delayed or even prevented by magnetic fields, radiative losses, morphology, and other factors. In addition, new clouds can be continuously injected through ejections from the outer accretion disk \citep[e.g.][]{Czerny2011}. A full discussion is beyond the scope of this paper, but the reader can turn up to the papers by \cite{Cooper2009} and \cite{Banda-Berragan2019} for simulations and discussions.}. The time it takes for a bowshock to form, $t_{\rm bw}$, is very short in comparison with all other relevant timescales, so we assume the bowshocks are in a steady state most of the time.\par

At the distances from the black hole where the interactions occur, the thermal cooling lengthscale is longer than the size of the bowshocks, and therefore shocks in the wind are adiabatic. On the contrary, shocks in the clouds are radiative. Only the former are efficient for particle acceleration, so we shall consider them in the estimates presented below.\par
\subsection{Nonthermal radiation}
\label{subsect:radiation}
We now calculate the nonthermal radiation emitted in the bowshock of a single cloud. The parameters adopted for the calculation of particle cooling and acceleration, distribution function in phase space, radiative emissivities, and fluxes are listed in Table \ref{table:radiation}.\par
We first estimate the maximum energy for each population of particles assuming that the minimum energy is $E_{\rm min}^{\rm e, p} = 2m_{\rm e, p}c^2$. 
\begin{table}[h]
\centering                          
\caption{\small Parameters adopted for the calculation of nonthermal processes}             
\label{table:radiation}      

\begin{tabular}{l c} 
\hline 
Parameter & Value  \\
\hline          
   Distance to the source & $5\,{\rm Mpc}$\\
   Bowshock kinetic luminosity & $L_{\rm k,bs} = 2.3\times10^{31}\,{\rm erg\,s^{-1}}$\\
   Bowshock size & $Z = 9.5\times10^{10}\,{\rm cm}$\\
   Relativistic particles luminosity &  $L_{\rm rel} = 2.3\times10^{30}\,{\rm erg\,s^{-1}}$\\
   Particle minimum energy & $E_{\rm min}^{\rm e,p} = m_{\rm e,p}c^2$ \\
   Electron maximum energy & $E_{\rm max}^{\rm e} = 0.3\,{\rm TeV}$ \\
   Proton maximum energy & $E_{\rm max}^{\rm p} = 37.8\,{\rm TeV}$\\
   Hadron-to-lepton power ratio & $a = 0.01,\,100$ \\
   Magnetization parameter  & $\alpha_{\rm B} = 10^{-1}$\\ 
   Magnetic field in the bowshock & $B_{\rm bs} = 28.8\,{\rm G}$ \\
   Wind density in the bowshock & $n_{\rm bs} = 10^8\,{\rm cm^{-3}}$ \\
   Injection spectral index & $p = 2$ \\
   Diffusion coefficient& $D = D_{\rm Bohm}$ \\
   \hline    
\end{tabular}
\end{table}
\begin{figure*}[h!]
 \centering
  \subfloat[\small electrons]{
   \label{fig:wind temperature 2}
    \includegraphics[scale=0.6]{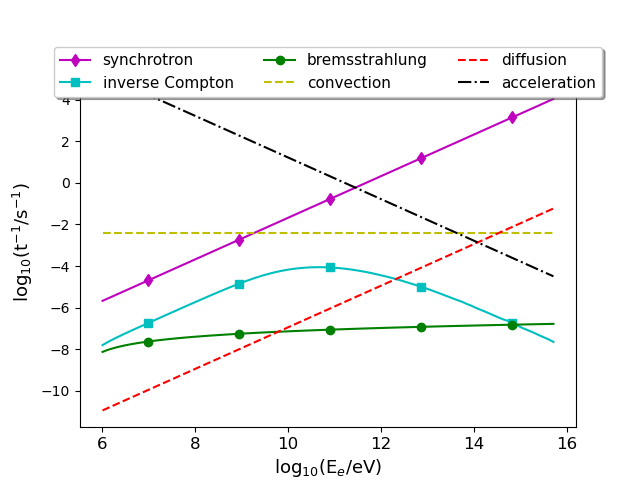}}
  \subfloat[\small protons]{
   \label{fig:wind SED 2}
    \includegraphics[scale=0.6]{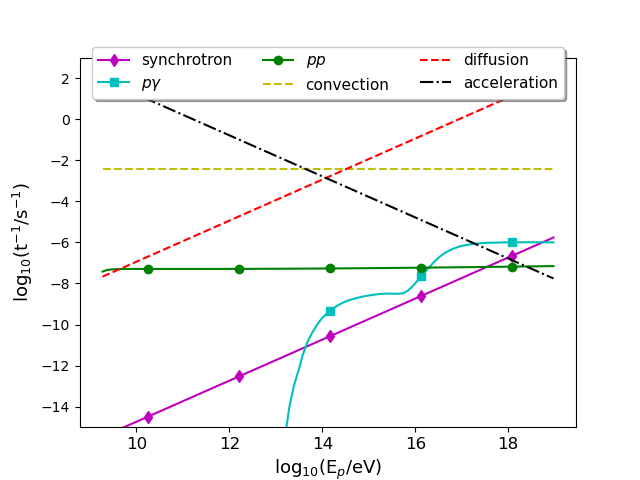}}
 \caption{\small Acceleration, cooling and escape timescales for relativistic electrons and protons. We assume diffusion in the Bohm regime.}
 \label{fig:particles timescales}
\end{figure*}

\begin{figure}
\centering
  \begin{minipage}{0.7\textwidth}
    \includegraphics[width=0.7\textwidth]{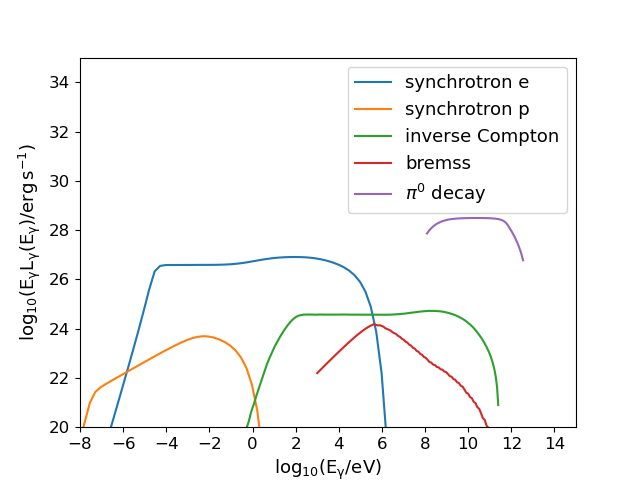}\\
   \end{minipage}%
  \hspace{5mm}
  \begin{minipage}{0.7\textwidth}
    \includegraphics[width=0.7\textwidth]{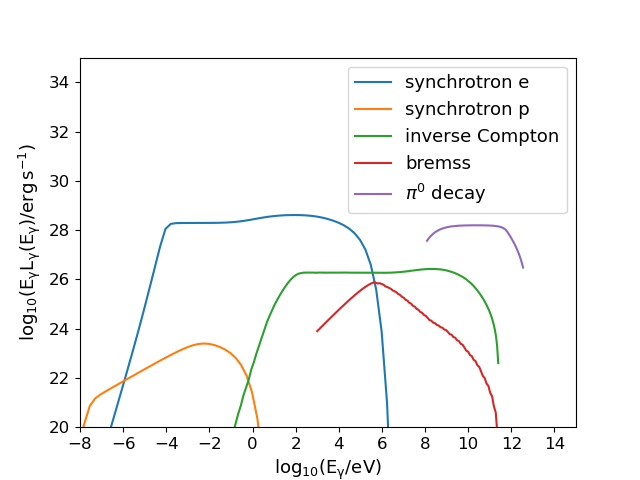}\\
  \end{minipage}
  \hspace{5mm}
  \begin{minipage}{0.7\textwidth}
    \includegraphics[width=0.7\textwidth]{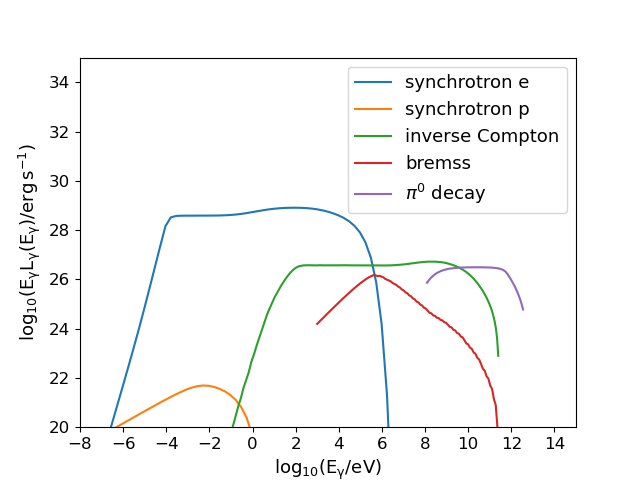}\\
  \end{minipage} 
  \caption{\small nonthermal SEDs for different choices of hadron-to-lepton} power ratio. Top: $a = 100$, medium: $a = 1$, bottom: $a = 0.01$. In all cases the adopted minimum energy is $E_{\rm min,\,e,p} = 2m_{\rm e,p}c^2$, and the index spectral of particle injection is $p=2$. Absorption effects are not relevant in any case. The radiation emitted in a single bowshock is shown.
  \label{fig:comparation SEDs}
\end{figure}

\begin{figure*}[h]
 \centering
      \subfloat[\small $a = 100$]{
      \label{fig:SED_comparation5}
      \includegraphics[scale=0.6]{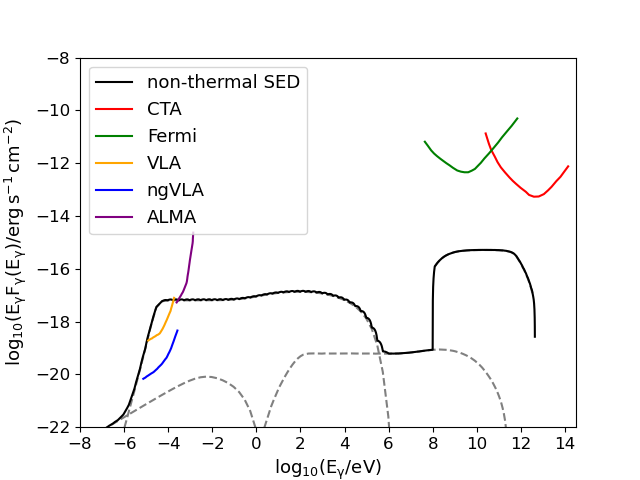}}
      \subfloat[\small $a = 0.01$]{
      \label{fig:SED_comparation6}
      \includegraphics[scale=0.6]{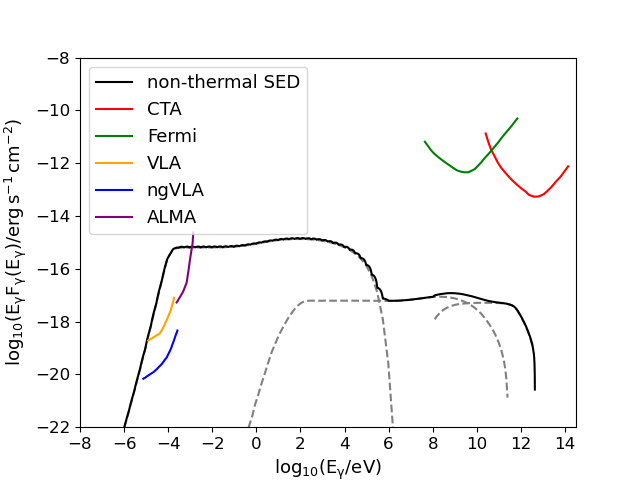}}
      \caption{\small Predicted flux of the nonthermal emission from wind-cloud interactions for an AGN located at $d = 5\,{\rm Mpc}$ respect to the observer. We assume $N_{\rm c}^{\rm w} = 5\times10^7$ interactions with identical clouds at $r = 10^{16}\,{\rm cm}$  from the supermassive black hole.}
      \label{fig:comparation flux at different a}
\end{figure*}

Figure \ref{fig:particles timescales} shows the timescales for the energy gain and losses of the relativistic particles. We consider synchrotron radiation, inverse Compton scattering of electrons with the radiation field, and gamma-ray production by the decay of neutral pions $\pi^0 \rightarrow \gamma+\gamma$ generated in inelastic $pp$ collisions. We assume that the magnetic energy density in the wind is a small factor of the kinetic energy density; hence, the hydrodynamical model of the wind remains valid. We quantify this ratio with a magnetization parameter $\alpha_{\rm B} = u_{\rm mag}/u_{\rm K}$, where $u_{\rm mag}$ is the magnetic energy density and $u_{\rm K}$ is the kinetic energy density. Considering amplification effects of the magnetic field by instabilities in the bowshock, we adopt a fiducial value $\alpha_{\rm B} = 10^{-1}$ \citep[see e.g.][]{romero2018b}. This leads to a magnetic field $B \approx 30\,{\rm G}$, consistent with the values derived by \cite{piotrovich2017} for broad-line regions from polarimetric observations. The wind photosphere provides the target photons for inverse Compton. We assume that the particles are isotropically distributed over pitch angles and that the medium is locally homogeneous. We calculate the escape time by advection and diffusion of the particles assuming the Bohm regime. We all this assumptions we obtain:
\begin{equation}
E_{\rm max}^{\rm e} = 3.0\times10^{11}\,{\rm eV},
\end{equation}
and
\begin{equation}
E_{\rm max}^{\rm p} = 3.8\times10^{13}\,{\rm eV}.
\end{equation}

The size of the acceleration region is a constraint on the maximum energy because it cannot exceed the particle gyroradius $r_{\mathrm{g}}= E / eB$ (Hillas criterion). The maximum energy of the electrons is determined by synchrotron losses, whereas the escape is the dominant loss for protons. These energies are well below the Hillas limit:
\begin{equation}
E_{\rm max}^{\rm Hillas} = eB_{\rm bs}Z \approx 8.0\times10^{14}\,{\rm eV}.
\end{equation}

The kinetic power of a single bowshock is 
$L_{\rm k,bs} = 2.3\times10^{31}\,{\rm erg\,s^{-1}}$. We assume that 10\% of this power is transformed into relativistic particles. Than we analyze three scenarios for the distribution of the energy between electrons and protons: $a = 100,\,1,\,{\rm and}\,0.01$.\par

We compute the particle distribution by solving Eq. (\ref{eq:particle transport}). 
In Fig. \ref{fig:comparation SEDs} we show the SEDs for the interaction of a single cloud with the wind for the different relative powers between protons and electrons. Total radiation power of the bowshocks range from $10^{28}$ to a few times $10^{29}$ erg s$^{-1}$, so for $5\times 10^7$ interacting clouds with have nonthermal bolometric luminosities of $\sim 10^{35-36}$ erg s$^{-1}$.  

In all cases synchrotron radiation dominates the spectrum from radio up to hard X-rays, inverse Compton scattering is the most important radiative process at high energies up to GeV, and gamma rays by neutral pion decays produce the emission in the TeV range. 


We finally compute the flux observed on Earth considering multiple interactions by many identical clouds. We assume that radiation from all clouds inside the wind contributes to the total flux. This is valid because most of the clouds are confined to distances from the black hole of $r\sim R_{\rm BLR}$, very far from the photosphere. Some clouds are eventually destroyed within the wind, but are replaced by new clouds so we assume that, on average, their number is roughly constant. 

In order to offer some figures, let us consider an AGN located at $d = 5\,{\rm Mpc}$. In Fig. \ref{fig:comparation flux at different a} we show the flux that would be observed on Earth for $N_{\rm c}^{\rm w}=5\times10^7$ identical clouds, all interacting at the same distance from the compact object. We see that, regardless of the hadron-to-lepton power ratio, the AGN can be  detected at radio frequencies with current instrumentation. At very-high energies, on the contrary, the emission is not strong enough for detection.\par



\section{Application to Mrk 335}
\label{sect:application}
We apply our model to the super-accreting AGN Mrk 335. This AGN is classified as a Narrow-Line Seyfert 1 galaxy. It harbors a supermassive black hole of $M_{\rm BH} \approx \left(1-3\right)\times10^7\,M_{\odot}$ \citep[see e.g.][and references therein]{kassebaum1997,peterson1998,peterson2004,zu2011,grier2012,yao2021}. 
This source has been extensively investigated in radio \citep{yang2020,yao2021}, infrared \citep{martinez-paredes2017}, optical \citep{doroshenko2005}, X-rays \citep{parker2014,wilkins2015,mondal2021}, and gamma rays \citep[not detected, just upper limit established, see][]{kataoka2011}. The geometry and kinematics of the broad-line region in Mrk 335 have been investigated in \cite{grier2017}. They found that Mrk 335 is best described by a thick disk that is close to face-on to the observer. The kinematics of BLR clouds consists of elliptical orbits of very low eccentricity.\par

The radiation escaping from the wind photosphere depends on the viewing angle because the surface defined by $\tau_{\rm ph}=1$ is not exactly spherical but slightly concave in the innermost region. Hence, for small inclinations of the normal to the disk with the line of sight, X-rays from the disk can escape directly to the observer \citep[see][]{fukue2009}. This effect explain why some X-rays fluxes around $10^{-11}$ erg s$^{-1}$ cm$^{-2}$ have been detected from Mkr 335 with \textit{Suzaku}, XMM-Newton and \textit{NuSTAR} \citep[see][and references therein]{keek2016}. The inclination angle of the galaxy is smaller than 30 - 45 degrees \citep{grier2017}. 

The bolometric luminosity of Mrk 335 is $L_{\rm bol} = 10^{44.9-45.7}\,{\rm erg\,s^{-1}}$ \citep{castello-mor2016}, which implies an Eddington ratio $\lambda_{\rm Edd} = 0.48-3.11$ \citep{yang2020}. For such a ratio we estimate using Eq. (\ref{eq:disk luminosity}) a range of possible values for the accretion rate of
\begin{equation}
\dot{m} \approx 1.3 - 1200.
\end{equation}
In order to make some quantitative estimates we adopt a hypercritical accretion rate of $\dot{m} = 1000$. The mass-loss rate in the wind is approximately the same.\par

The radio flux of Mrk 335 has been measured at different levels of resolution. Recently, \cite{yang2020} reported the flux at three radio frequencies: 1.4, 5, and 8.4 GHz, with spatial resolutions of 1.24, 0.204, and 0.154 kpc, respectively. The spectral index (logarithmic slope $\alpha$, where $F_{\nu} \propto \nu^{\alpha}$) between 5 and 8.4 GHz was estimated as $\alpha_{\rm r} = -0.84$, which does not favor an origin of the radio emission from synchrotron radiation in a jet (which is expected to be flatter).  \cite{yao2021}, using VLBA images at 1.5 GHz, reported the detection of parsec-scale bipolar outflows at Mrk 335. They suggest that the associated brightness temperature might be indicative of nonthermal radiation. The measured total flux density is similar to the value reported by \cite{yang2020} at 1.4 GHz, which indicates that the radio emission within 1 kpc is dominated by nuclear component.\par

\begin{figure}[h!]
 \centering 
 \includegraphics[scale=0.6]{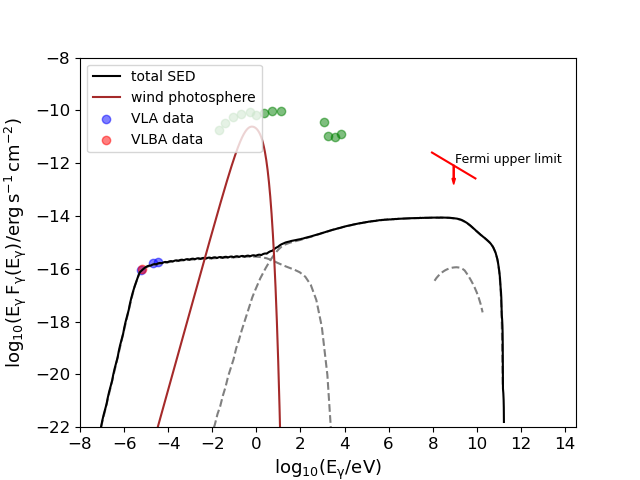}
 \caption{\small Nonthermal spectrum of Mrk335. We show VLA and VLBA data obtained by \cite{yang2020} and \cite{yao2021}. he dashed lines correspond to individual SEDs: electron synchrotron, inverse Compton scattering, and $pp$ collisions. The green points represent the archival data from NED (see http://nedwww.ipac.caltech.edu), and they come from other regions of the source. Fitting parameters are listed in Table \ref{table:mrk335}.}
 \label{fig:SED mrk335}
\end{figure}

We adopt a magnetization parameter $\alpha_{\rm B} = 10^{-5}$ that corresponds to a magnetic field in the clouds of of $B_{\rm bs} = 0.1\,{\rm G}$. This leads to inverse Compton scattering emission dominating the spectrum at high energies and adequately fits the observed radio emission as synchrotron radiation by relativistic electrons produced in wind-cloud interactions close to the black hole. This allows for steeper radio spectral indices. In Fig. \ref{fig:SED mrk335} we show the fit of pour model to the VLA and VLBA data obtained by \cite{yang2020} and \cite{yao2021}. We adopt a distance to the source of $d = 86\,{\rm Mpc}$ \citep{wang2014}. The number of BLR clouds and the filling factor are $N_{\rm c} = 10^{12}$ and $f_{\rm BLR} = 2\times10^{-3}$. Notice that these values are higher than those considered in section \ref{sect:results}. The large number of clouds could be the result of clump formation in the wind as shown, for instance in the 2D radiation-hydrodynamic by 
\cite{takeuchi2013}. Regarding the filling factor, it can be calculated as in  \cite{takeuchi2013}, see their Eq. (25). 

Blackbody radiation from the wind photosphere contributes to the flux observed at 0.1 - 1 eV, where the dominant components are expected to be the outer disk and a thick torus around the central black hole. 

\begin{table}[h]
\centering                          
\caption{\small Model-fitting parameters of VLA and VLBA data of Mrk335.}             
\label{table:mrk335}      

\begin{tabular}{l c} 
\hline 
Parameter & Value  \\
\hline       
   Distance & $d = 86\,{\rm Mpc}$ \\   
   Wind mass-loss rate & $\dot{M}_{\rm w} = 1000\,\dot{M}_{\rm cr}$\\
   Black hole mass & $M_{\rm BH} = 10^{7}\,M_{\odot}$\\
   Broad-line region size & $R_{\rm BLR} = 4\times10^{4}\,r_{\rm g}$\\   
   Number of BLR clouds &  $N_{\rm c} = 10^{12}$\\
   Broad-line region filling factor & $f_{\rm BLR} = 2\times10^{-3}$ \\
   Relativistic particles fraction &  $q_{\rm rel} = 10^{-1}$\\   
   Magnetization parameter  & $\alpha_{\rm B} = 10^{-5}$\\ 
   Hadron-to-lepton energy ratio & $a = 0.01$\\
   Injection spectral index & $p = 2.0$ \\
   Diffusion coefficient& $D = D_{\rm Bohm}$ \\
   \hline    
\end{tabular}
\end{table}
The parameters we use to fit the radio data are listed in Table \ref{table:mrk335}. Our results, as seen from Fig. \ref{fig:SED mrk335}, can explain the nuclear radio emission without invoking a jet. The spectrum results naturally from standard injection and Bohm diffusion in the interaction zone.

\section{Discussion}
\label{sect:discussion}
\subsection{Sensitivity to model parameters}
\label{subsect:sensitivity}
In Sect. \ref{sect:results}, we have adopted specific parameters to investigate wind power, cloud hydrodynamics, and nonthermal emission in bowshocks. Here we discuss how our results change with the variation of the parameters.\par

The accretion rate determines the wind terminal velocity and the location of the photosphere through Eqs. (\ref{eq:wind velocity}) and (\ref{eq:apparent photosphere}). If the accretion rate is $\dot{m} < 100$, the wind is faster and the photosphere is located closer to the central black hole. The wind number density depends on the accretion rate as $n_{\rm w} \propto \dot{M}^{3/2}$, therefore the accretion rate regulates the dynamics of BLR clouds. For example, for $\dot{m} = 10$, the region where BLR clouds can orbit without being destroyed before fully entering the wind is almost twice as large than in the case of $\dot{m}=100$. The maximum energy of the relativistic particles for the smaller accretion rate is slightly increased, and the SED practically does not change. Inverse Compton radiation is less relevant because the radiation field in the interaction region is more diluted. If the accretion rate, on the contrary, is very high, e.g. $\dot{m} = 1000$, the wind photosphere is very close to the edge of the broad-line region $R_{\rm ph}\approx 0.9R_{\rm BLR}$. The total SED changes slightly, with an increase in inverse Compton radiation. In this scenario, fewer clouds contribute to the total flux of the source and the overall luminosity is smaller by a factor $N_{\rm c}^{\rm w\,\prime}/N_{\rm c}^{\rm w}$, where $N_{\rm c}^{\rm w\,\prime}$ is the number of clous inside the wind for $\dot{m}=1000$.\par

The magnetic field in the bowshocks is assumed to be weak enough for the hydrodynamic approximation to be valid. Magnetic fields can attenuate hydrodynamic instabilities and increase cloud lifetime  \citep{shin2008}. 
If the magnetic field is stronger, e.g. $\alpha_{\rm B} = 0.9$, the maximum energy of electrons decreases to $E_{\rm max}^{\rm e} \approx 0.2\,{\rm TeV}$, and for protons it increases approximately by a factor of three $E_{\rm max}^{\rm p} \approx 0.1\,{\rm PeV}$. 
If the magnetic field is very weak such that $\alpha_{\rm B} = 10^{-6}$, the maximum energy of protons and electrons is set by convection to $E_{\rm max} = 0.1\,{\rm TeV}$ for both species. In this scenario, the hadronic emission is almost totally suppressed but a leptonic model remains viable. The synchrotron spectrum decreases significantly, and the inverse Compton scattering is the dominant contribution.\par

Typically, $f_{\rm BLR} = 10^{-6}$ is adopted for the filling factor in BLRs. In the presence of a disk wind, however, this value changes mainly because of two reasons: 1) the clouds heat up and expand, hence the size of the bowshocks grows; 2) at scales beyond the photosphere, the wind adopts a clumpy structure, so in addition to the BLR clouds the contribution of the over-densities of the wind itself must be added \citep[see][]{takeuchi2013,kobayashi2018}. Assuming that, when considering both effects, the filling factor increases to $f_{\rm BLR} = 10^{-4}$, the available kinetic power in a single bowshock also increases by two orders of magnitude, and the synchrotron self-absorption cut-off shifts slightly to higher energies.\par


\subsection{Distribution of BLR clouds in the wind}
We assumed in Sect. \ref{sect:results} that all the clouds orbit at the same distance from the black hole. A more realistic estimate of the total power available in nonthermal particles is obtained by assuming that the clouds orbit within a spherical shell of finite thickness,. The interacting clouds are those within the solid angle subtended by the wind. \par

We assume that the shell with the clouds has an inner radius $R_{\rm in}$ and an outer radius $R_{\rm BLR}$. The total number of BLR clouds is given by Eq. (\ref{eq:filling factor}), and the number of BLR clouds inside the wind by Eq. (\ref{eq:clouds inside the wind}). The total power available in all bowshocks can be estimated as
\begin{equation}
L_{\rm bs}^{\rm tot} = N_{\rm c}^{\rm w} \langle L_{\rm bs}^1 (r) \rangle,
\end{equation}
where $L_{\rm bs}^1 (r)$ is the kinetic power injected into a single bowshock at $r$, and $\langle L_{\rm bs}^1 (r)\rangle$ is the average value given by

\begin{equation}
\label{eq:average single power}
\langle L_{\rm bs}^1 \rangle = \frac{\int_{R_{\rm in}}^{R_{\rm BLR}} {\rm d}r \frac{{\rm d}N_{\rm c}^{\rm w}}{{\rm d}r}L_{\rm bs}^{1}(r)}
{{\int_{R_{\rm in}}^{R_{\rm BLR}} {\rm d}r \frac{{\rm d}N_{\rm c}^{\rm w}}{{\rm d}r}}},
\end{equation}
where ${\rm d}N_{\rm c}^{\rm w}$ is the number of clouds in a differential volume ${\rm d}V_{\rm w} = \Omega_{\rm w}r^2 {\rm d}r$. Assuming that the clouds are uniformly distributed, 
the total power injected into nonthermal particles results
\begin{equation}
L_{\rm rel}^{\rm tot} = \frac{3}{2}q_{\rm rel} f_{\rm BLR}\left(1 + \frac{Z}{R_{\rm c}} \right)^3 \frac{L_{\rm k,w}}{v_{\rm w}^3} \frac{\int_{R_{\rm in}}^{R_{\rm BLR}} {\rm d}r\, r^2\,v_{\rm sw}^3 (r)}{\left(R_{\rm BLR}^3 - R_{\rm in} ^3\right)/3},
\end{equation}
where we assume that the fraction of the kinetic power transformed into relativistic particles $q_{\rm rel}$ is the same in each bowshock. Typically, the clouds are much denser than the wind, so we can adopt the asymptotic value of $v_{\rm sw}$. Then,
\begin{equation}
L_{\rm rel}^{\rm tot} \approx \frac{3}{4}q_{\rm rel} f_{\rm BLR} \left(1 + \frac{Z}{R_{\rm c}} \right)^3 \left(1 + \frac{v_{\rm c}^2}{v_{\rm w}^2} \right)^{3/2} \dot{M}_{\rm w}v_{\rm w}^2.
\end{equation}

Assuming fiducial values $Z = 0.3\,R_{\rm c}$, and $v_{\rm w} = 5\,v_{\rm c}$, we obtain $L_{\rm rel}^{\rm tot} \approx 1.75 q_{\rm rel} f_{\rm BLR} \dot{M}_{\rm w}v_{\rm w}^2$.

If we adopt $q_{\rm rel} = 0.1$, $\dot{M}_{\rm w} = 100\,\dot{M}_{\rm cr}$ and $M_{\rm BH} = 10^7\,M_{\odot}$, we get $L_{\rm rel}^{\rm tot} \approx 5.5\times 10^{37}\,{\rm erg\,s^{-1}}$. The total power in nonthermal electromagnetic radiation is of the order of $10^{36}\,{\rm erg\,s^{-1}}$. This is not too far from what was estimated in Sect. \ref{sect:results}, so our first approximation is good within the order of magnitude.





\subsection{Variability}
The total radiative emissivity depends on the number of clouds inside the wind, which is expected to vary randomly around a constant value given by $\langle N_{\rm c}^{\rm w} \rangle =  N_{\rm c} \Omega_{\rm w}/2\pi$, if cloud destruction within the wind is of the order of cloud injection. The radiation caused by the interactions between the clouds and the wind turns on and off with the injection and disappearance of the individual clouds.  The random fluctuations of $N_{\rm c}^{\rm w}$ imprint a noise-like variability in the lightcurves of the whole source. The number of clouds within the wind changes by the permanent entry and exit of clouds, as well as by the destruction of the clouds by hydrodynamic instabilities and by the effect of shocks in the clouds. We make below a simple estimate of the relative variation of the nonthermal luminosity emitted from these sources adopting the model developed by \cite{owocki2009}.\par

Defining the porosity length as $h = R_{\rm c}/f_{\rm BLR}$, from energy considerations and assuming standard statistics, the relative fluctuation in luminosity is given by Eq. (5) in \cite{owocki2009}:
\begin{equation}
\frac{\delta L}{\langle L \rangle} = \sqrt{\frac{h}{\pi R_{\rm BLR}}} \sim \frac{1}{\sqrt[6]{N_{\rm c}\,f_{\rm BLR}^2}}.
\end{equation}

Small flickering is expected only for a high number of clouds or a high volume filling factor. This could be the case if a negligible number of clouds are removed by wind flow, or if the filling factor is raised by some mechanism such as those discussed at the end of subsect. \ref{subsect:sensitivity}. Assuming $N_{\rm c} = 10^{9}$ and $f_{\rm BLR} = 5\times10^{-4}$, we obtain $\delta L/L \approx 0.11$, which corresponds to a flickering level of $\approx 11\%$. This occurs on a timescale of less than 2 hours.

\section{Conclusions}
\label{sect:conclusions}
We have developed a semi-analytical model for the nonthermal emission from the central region of super-accreting AGNs. The model is based on the interaction of strong winds driven by radiation from the disk with clouds of the broad line region. The model predicts detectable radio emission in AGNs without the need to assume a low-power jet. This emission is synchrotron radiation produced by relativistic electrons accelerated in the bowshocks that form around the clouds in their motion within the wind. \par
Only radiation produced outside the photosphere of the wind can escape from the source. Our model, then, is adequate to account for radio emission in super-accreting AGNs, but not hyperaccreting ones. Roughly, the range of applicability is  for objects with accretion rates of $\dot{m} \leq 10^4$, if the location of the BLR is $R_{\rm RBL} \sim  10^{4}\,r_{\rm g}$.\par
Our model is easily adaptable to any super-Eddignton source surrounded by clumps of matter. So, it is natural to apply it to high-mass X-ray binaries, transient sources as those created by tidal disruption events, and ultraluminous X-ray sources. In the case of X-ray binaries and super-accreting microquasars, the wind from the massive star has a clumpy structure, which permeates all of space at the scales of the binary system. The wind from the disk launched in super-critical accretion episodes should interact with the clumps producing a similar phenomenology to the one discussed in this paper, but at a quite different scale. The detailed radiative output will be explored in a future communication.

\begin{acknowledgements}
The authors thank the anonymous referee for a careful and constructive review, and for his/her comments that improved this work. PS thanks Leandro Abaroa for the fruitful discussions about this research. PS and GER acknowledge support by PIP 2021-1639 (CONICET). GER acknowledges the support by the Spanish Ministerio de Ciencia e Innovación (MICINN) under grant PID2019-105510GB-C31 and through the “Center of Excellence María de Maeztu 2020-2023” award to the ICCUB (CEX2019-000918-M).
\end{acknowledgements}
%
%

\bibliographystyle{aa} 
\bibliography{biblio}

\begin{appendix}
\section{Timescales for relativistic particles}\label{app1}

\subsection{Cooling}
Radiative losses are caused by the interaction of the particles with environmental fields. The cooling timescale for a charged particle of energy $E$ and mass $m$ by synchrotron radiation is \citep[e.g.][]{blumenthal1970}:
\begin{equation}
t_{\mathrm{synchr}}^{-1} = \frac{4}{3}\left(\frac{m_{\mathrm{e}}}{m}\right)^{2}c\sigma_{\mathrm{T}}u_{\mathrm{B}}\frac{E}{m^{2}c^{4}},
\label{equation_SynchrotronCooling}
\end{equation}
where $u_{\mathrm{B}}=B^{2}/8\pi$ is the magnetic energy density.\par
Relativistic electrons interacting with radiation fields produce high-energy radiation by inverse Compton scattering. The cooling timescale for an electron of energy $E$ in a target radiation field is:
\begin{equation}
t_{\mathrm{IC}}^{-1} = \frac{1}{E}\int_{E_{\mathrm{ph}}^{\mathrm{min}}}^{E_{\mathrm{ph}}^{\mathrm{max}}} \mathrm{d}E_{\mathrm{ph}} \int_{E_{\mathrm{ph}}}^{E_{\gamma}^{\mathrm{max}}} \mathrm{d}E_{\gamma} (E_{\gamma} - E_{\mathrm{ph}}) P_{\mathrm{IC}}(E,E_{\mathrm{ph}},E_{\gamma}),
\label{equation_iComptonCooling}
\end{equation}
where the function $P_\mathrm{IC}(E,E_{\mathrm{ph}},E_{\gamma})$ is given in \cite{blumenthal1970}.\par
Relativistic protons interacting with radiation fields can produce pairs and pions. The cooling timescale by the production of these particles for a relativistic proton of energy $E$ is given by \citep{stecker1968}:
\begin{equation}
\begin{split} 
t_{p\gamma}^{-1} =&\frac{m_{\mathrm{p}}^{2}c^{5}}{2E^{2}}\int_{\frac{E_{\mathrm{ph(thr)}}^{\prime(i)}}{2\gamma_{p}}}^{E_{\mathrm{ph}}^{\mathrm{max}}}\mathrm{d}E_{\mathrm{ph}}\frac{n_{\mathrm{ph}}(E_{\mathrm{ph}})}{E_{\mathrm{ph}}^{2}} \times \\
&\int_{E_{\mathrm{ph(thr)}}^{\prime(i)}}^{2\gamma_{\mathrm{p}}E_{\mathrm{ph}}}\mathrm{d}E_{\mathrm{ph}}^{\prime}{}^{\prime}\sigma_{p\gamma}^{i}(E_{\mathrm{ph}}^{\prime})K_{p\gamma}^{i}(E_{\mathrm{ph}}^{\prime})E_{\mathrm{ph}}^{\prime},
\end{split} 
\end{equation}
where $E_{\mathrm{ph(thr)}}^{\prime}$ is the photon threshold energy for the pairs creation ($i=e^{\pm}$) or a pions production ($i=\pi$).\par
The interaction of relativistic electrons with the cold proton plasma produces relativistic Bremsstrahlung radiation. The timescale cooling for an electron of energy $E$ is \citep{berezinskii1990}:
\begin{equation}
t_{\mathrm{Br}}^{-1}=4\alpha_{\mathrm{FS}}r_{\mathrm{e}}^{2}Z^{2}cn_{\mathrm{p}}\left[\ln\left(\frac{2E}{m_{\mathrm{e}}c^{2}}\right)-\frac{1}{3}\right],
\label{equation_bremssCooling}
\end{equation}
where $\alpha_{\mathrm{FS}}$ is the fine structure constant and $n_{\mathrm{p}}$ is the cold proton density.\par
Relativistic protons can also interact with the cold protons. If the energy of the relativistic proton is higher than $1.22\,\mathrm{GeV}$ the collision with a cold proton can create pions. The cooling timescale for a proton of energy $E$ is given by \citep[see][]{begelman1990}:
\begin{equation}
t_{pp}^{-1} \approx cn_{\mathrm{p}}K_{pp}\sigma_{pp}(E),
\end{equation}
where $\sigma_{pp}$ is the cross section of this process and $K_{pp}\approx 0.5$ is the inelasticity of the interaction.\par
Total radiative losses for relativistic electrons are calculated as:
\begin{equation}
t_{\mathrm{cool}}^{-1} = t_{\mathrm{synchr}}^{-1} + t_{\mathrm{IC}}^{-1} + t_{\mathrm{Br}}^{-1},
\end{equation}
whereas for relativistic protons they are:
\begin{equation}
t_{\mathrm{cool}}^{-1} = t_{\mathrm{synchr}}^{-1} + t_{p\gamma, e^{\pm}}^{-1} + t_{p\gamma, \pi}^{-1} + t_{pp}^{-1}.
\end{equation}

\subsection{Escape}
We consider two mechanisms for the escape of particles from the acceleration region: convection by fluid motion, and diffusion by multiple scattering. Particle escape in the context of wind-cloud interaction in starburst galaxies is detailed in \citet{muller2020b}.\par
The timescale of particle removal by convection is given by:
\begin{equation}
t_{\rm conv} = \frac{4~R_{\rm c}}{v_{\rm sw}},
\end{equation}
where the factor $4$ takes into account the effect of a strong shock. Diffusion by random particle motion occurs on a timescale given by:
\begin{equation}
t_{\rm diff} = \frac{R_{\rm c}^2}{D(E)},
\end{equation}
where $D(E)$ is the diffusion coefficient. We adopt the Bohm regime for particle diffusion; then
\begin{equation}
D(E) = D_{\rm B}(E) = \frac{r_{\rm g}c}{3},
\end{equation}
where $r_{\rm g} = E/eB$ is the particle gyroradius.
\subsection{Acceleration}
In this work, we consider only the diffusive shock acceleration. The particles gain energy in successive collisions with magnetic inhomogeneities at each side of the shock. The timescale for the acceleration is given by \citep[e.g.][]{aharonian2004}:
\begin{equation}
t_{\rm acc}^{-1}(E) = \eta^{-1} \frac{ecB}{E},
\end{equation}
where $\eta$ is the acceleration efficiency ($\eta \leq 1$) and its inverse is
\begin{equation}
\eta^{-1} = 20 \frac{D}{r_{\rm g}c}\left(\frac{c}{v_{\rm sw}}\right)^2,
\end{equation}
where $D$ is the diffusion coefficient. We assume Bohm diffusion.
\end{appendix}

\end{document}